\documentclass[aps,preprint,showpacs,superscriptaddress,groupedaddress]{revtex4}  
\usepackage{graphicx}  
\usepackage{dcolumn}   
\usepackage{bm}        
\usepackage{amssymb}   
\usepackage{subfigure}
\begin{document}

\widetext


\title{Periodic boundary conditions for the simulation of uniaxial extensional flow}
\author{Thomas A. Hunt}
\affiliation{Computational Biophysics, University of Twente, P.O. Box 217, 7500 AE Enschede, The Netherlands}
\email{tomahunt@gmail.com}

\date{October 15, 2013}

\begin{abstract}
It is very common with molecular dynamics and other simulation techniques to apply Lees-Edwards periodic boundary conditions (PBCs) for the simulation of shear flow. 
 However the behavior of a complex liquid can be quite different under extensional flow. 
 Simple deformation of a simulation cell and its periodic images only allows for simulations of these flows with short duration. 
 For the simulation of planar extensional flow it was recognized that the PBCs of Kraynik and Reinelt [Int. J. Multiphase Flow \textbf{18}, 1045 (1992)] 
 could be used to perform simulations of this flow with arbitrary duration.
 However, a very common extensional flow in industrial applications and experiment is uniaxial extensional flow.
 Kraynik and Reinelt found that their method could not be directly generalized to this flow because of the lack of a lattice which reproduces itself during uniaxial extension. 
 PBCs are presented in this article which solve this problem, by finding a lattice which is compatible with the flow, finding the reduced basis to the lattice at all times and using this basis when calculating the position and separation of particles.
 Using these new PBCs we perform nonequilibrium molecular dynamics simulations of a simple  liquid and show that the technique gives results which agree with those from simulations using simply deforming PBCs.  

\end{abstract}

\pacs{}
\maketitle

In an extensional flow fluid deforms in such a way that it is stretched in at least one direction and compressed in at least one other direction. Understanding the behavior of complex fluids under extensional flow is of particular importance to their application in industry, where, for example, their extrusion, fiber spinning and film blowing 
are affected considerably by their extensional behavior \cite{Mac94}.  Measuring properties of fluids under this type of flow is also particularly difficult and 
consequently techniques to simulate fluids at a microscopic scale under extensional flows have
 considerable value. 

 It is common to perform simulations of a fluid with 
periodic boundary conditions (PBCs) to obtain accurate bulk properties of the fluid both at 
equilibrium and out of equilibrium. There are several ways of describing PBCs 
mathematically and they can be applied to a very wide range of models. 
However, in this article we will take a pragmatic approach and describe them 
in a way common in the molecular dynamics literature.

When PBCs are applied \cite{AT87} one keeps a primary cell of particles 
and calculates the forces on the particles as if there were periodic images of the primary 
cell stacked around it (Fig. 1a). In this configuration a particle and its periodic images 
form a lattice. To perform simulations of flow one can use similar techniques to provide 
PBCs which change with the flow. For example for planar Couette flow 
one can use the Lees-Edwards PBCs \cite{LE72} or equivalently the Lagrangian-Rhomboid 
PBCs \cite{EM90}.

\begin{figure}[b!]
\label{fig:boxes}
\includegraphics[scale=1.0]{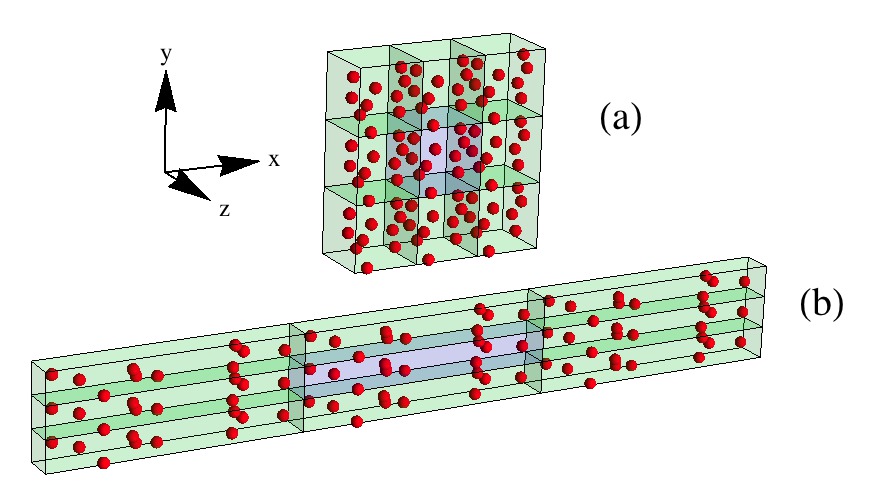}
\caption{\label{} (a) An equilibrium simulation cell (blue) and a plane of its periodic images (green) in the $xy$-plane. (b) The same simulation cell and periodic images deformed under uniaxial extensional flow to a Hencky strain $\epsilon=ln(4)$. After more extension particles interact with their periodic images.}
\end{figure}

When applying PBCs to a system under uniaxial extensional flow (UEF), where the velocity gradient of the fluid is given by $\nabla \mathbf{u} = \textrm{diag}(\dot\epsilon,-\dot\epsilon/2,-\dot\epsilon/2)$, the only existing technique is to take a simulation cell with periodic images and deform it in a way that is consistent with the flow: 
stretching in the $x$-direction and compressing in both the $y$ and $z$-directions. 
However, after some finite time the simulation cell becomes 
extremely long in the $x$-direction and narrow in both the $y$ and $z$-directions (Fig. 1b).
This means that in the $y$ and $z$-directions particles will interact with their own periodic images and the simulation fails. 
This technique has been applied by several authors to perform nonequilibrium molecular dynamics simulations (NEMD) of UEF up to a small extension \cite{Hey85,Evans90,Ryc90,Hou92}. 
We term these simple deforming PBCs.
When applied to a model of a complex fluid it may be that the relaxation time of the fluid is too long for the fluid to reach a steady state and so this technique is not sufficient to simulate these systems. 

In the case of planar extensional flow, with $\nabla \mathbf{u} = \textrm{diag}(\dot\epsilon,-\dot\epsilon,0)$, using simple deforming PBCs presents the same limitations. However, following a prior observation \cite{KH86} and the analysis of Adler and Brenner \cite{AB85}, Kraynik and Reinelt developed PBCs \cite{KR92} which avoided the problem. The technique was first applied to NEMD simulations by Todd and Daivis \cite{TDLett} and Baranyai and Cummings \cite{BC}. 
In summary, the lattice corresponding to a particle and its periodic images is rotated by a specific angle in such a way that a particle never comes closer than some fixed finite distance to its periodic images. This property is given the name \textit{compatibility}. In addition, after some Hencky strain $\epsilon_p = \tau_p \dot\epsilon$ the lattice is found to be mapped back onto its original configuration, termed \textit{reproducibility}. In effect, at $t=\tau_p$ the particles can be mapped back to their original cell and the simulation continued indefinitely. 
Kraynik and Reinelt show that no lattice exists which is reproduced under UEF and so an exactly equivalent technique cannot  be constructed for this flow. 
However, they comment that Adler \cite{Adler84} establishes the existence of lattices which are compatible under UEF and that strict compatibility may be established without the need for reproducibility. They also suggest that lattices might be found which are compatible for long but finite times. To the best of our knowledge, until now these points have not been utilized.

In this article we explicitly find a lattice which is strictly compatible under UEF and, in agreement with the analysis of  Kraynik and Reinelt, is not reproduced under the flow.  We then show how this lattice can be used to perform simulations of UEF up to a Hencky strain of at least $\dot\epsilon t = 1000$, the limit here being dependent on the numerical precision used during the calculation of the lattice. 

The remainder of the article is structured as follows. We begin by summarizing a well known technique for applying PBCs, thus simplifying the remaining analysis. We then find a lattice which is compatible under UEF. Following this a lattice reduction algorithm is developed which allows us to perform a mapping on the particles when required. With these elements a PBC algorithm is developed and summarized. In the last section the algorithm is applied to simulations of a simple liquid using nonequilibrium molecular dynamics.  

\section{Periodic boundary conditions}
In the previous section we described how PBCs can be applied by having a primary simulation cell surrounded by periodic images. We also noted that a particle and its periodic images form a lattice. It is well known that the Wigner-Seitz cell of the lattice could be used as the simulation cell \cite{Ber07}. However, there exists a very efficient
algorithm using the dual cell to the Wigner-Seitz cell \cite{Ber07,Smith89}. The dual cell is the \textit{parallelepiped} constructed from lattice vectors perpendicular to the faces of the Wigner-Seitz cell. These basis vectors form the most compact basis of the lattice and are termed the reduced basis vectors of the lattice.

The basis vectors of the lattice are conveniently represented by the rows of the matrix $\mathbf{B}$ termed the basis matrix.
As long as  the minimum distance between the faces of the dual cell is greater than twice the cut-off radius of the interatomic potential \cite{Smith89} the minimum image vector between particles can be calculated as follows \cite{Ber07}: the components are written in terms 
of the basis vectors of the lattice $\mathbf{r}_{ij} = \boldsymbol{\rho}_{ij}\cdot\mathbf{B} $ so that  $\boldsymbol{\rho}_{ij} = \mathbf{r}_{ij}\cdot\mathbf{B}^{-1}$; subtracting the integer part of $\boldsymbol{\rho}_{ij}$ we obtain the minimum image separation $\boldsymbol{\rho}_{ij}^{min} = \mathbf{r}_{ij}\cdot\mathbf{B}^{-1} -  \lfloor \boldsymbol{\rho}_{ij}\rceil$ and finally $\mathbf{r}^{min}_{ij} = \boldsymbol{\rho}_{ij}^{min}\cdot\mathbf{B}$.  The same transformations are also performed on the particles' centers of mass. The following definitions have been used: $\lfloor x\rceil=\lfloor x + 1/2 \rfloor$ is the nearest integer to $x$ and $\lfloor x \rfloor$ is the floor of $x$.  
\section{Compatibility}
\begin{figure}[t!]
\includegraphics[scale=0.4]{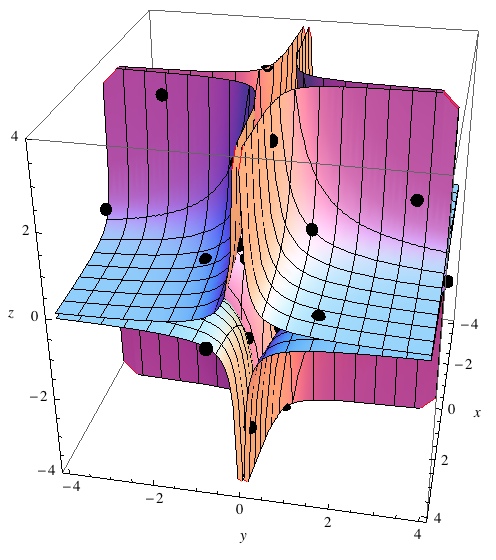} 
\caption{\label{fig:surface}The surface $|xyz|=1$ together with those points of the lattice $ n_1 \mathbf{b}_1 + n_2 \mathbf{b}_2 + n_3 \mathbf{b}_3$ which lie on the surface in the range of the plot.}
\end{figure}

During UEF the velocity gradient is given by $\nabla \mathbf{u} = \textrm{diag}(\dot{\epsilon},-\dot{\epsilon}/2,-\dot{\epsilon}/2)$. Under such a flow the fluid has a deformation,
$\mathbf{r}(t)=\mathbf{r}(0) \cdot e^{t\nabla\mathbf{u}}= (x(0)e^{\dot{\epsilon}t},y(0)e^{-\dot{\epsilon}t/2},z(0)e^{-\dot{\epsilon}t/2}) $ where $\dot\epsilon$ is termed the Hencky strain rate. We see from this expression that any point beginning on the surface $|xyz| = K$ (Fig. \ref{fig:surface})
will remain on that surface throughout the flow, while points outside this surface will not cross it during the flow. 
The closest that a point could come to the origin during the flow is $d_{min}=\sqrt{3}\sqrt[3]{K}$. A lattice which, except for the origin, lies on or outside this surface will be compatible with the flow.

If a primary particle sits at the origin then its periodic images form a lattice
\begin{equation}
\mathbf{r}_{\mathbf{n}}(t) = n_1 \mathbf{b}_1(t) + n_2 \mathbf{b}_2(t) + n_3 \mathbf{b}_3(t) 
\label{eq:lattice}
\end{equation}  
where $n_1, n_2$ and $n_3$ are integers and $\mathbf{b}_1(t),\mathbf{b}_2(t)$ and $\mathbf{b}_3(t)$ 
are the basis of the lattice at time $t$. If all points of the lattice except the origin lie on or outside the 
surface $|xyz|=K$ then during the flow all points will remain on or outside this surface. 

We now produce a lattice with this property for $K = 1$. Following an argument explained in Cassels \cite{Cass} and originally from Davenport \cite{Dav38}, we take the cubic 
equation
\begin{equation}
\phi^3 - 6\phi^2 + 5\phi - 1 = 0
\label{eq:cubic}
\end{equation}
which has three positive real solutions 
\begin{equation}
\phi_1 \approx 5.0489, \phi_2 \approx 0.6431, \phi_3 \approx 0.3080.
\label{eq:solns}
\end{equation}
From these solutions we construct the basis vectors, 
\begin{equation}
\mathbf{b}_1=(1,1,1),\mathbf{b}_2=(\phi_1,\phi_2,\phi_3),\mathbf{b}_3=(\phi_1^2,\phi_2^2,\phi_3^2). 
\label{eq:phi_basis}
\end{equation}
With this choice, lattice points have the components
\begin{eqnarray}
x_{\mathbf{n}}&=& n_1 + n_2 \phi_1 + n_3 \phi_1^2, \\
y_{\mathbf{n}}&=& n_1 + n_2 \phi_2 + n_3 \phi_2^2, \\
z_{\mathbf{n}}&=& n_1 + n_2 \phi_3 + n_3 \phi_3^2. 
\end{eqnarray} 
The product of the components is an \textit{integer valued} multinomial with ten terms:
\begin{equation}
|x_{\mathbf{n}}y_{\mathbf{n}}z_{\mathbf{n}}|= 
|n_1^3 + 6n_1^2 n_2 + \ldots  + 5 n_2 n_3^2 + n_3^3|.
\end{equation}
 If the product is zero then at least one of the three factors must 
 be zero. However, since these are quadratic polynomials in the solutions of an irreducible cubic equation the only 
 possible zero is with $n_1=n_2=n_3=0$.  The product $|x_{\mathbf{n}}y_{\mathbf{n}}z_{\mathbf{n}}|$ is therefore greater than or equal to $1$, except for the point at the origin, fulfilling our requirement. Any irreducible cubic with integer coefficients and real solutions could 
 have been used in this demonstration, however it will be seen that the properties of the matrix with row vectors $\mathbf{b}_i$ 
 will be used in the next section.  The lattice with basis (\ref{eq:phi_basis}) has $d_{min}=\sqrt{3}$.

\section{Lattice reduction}

 As mentioned above, to apply the PBCs we need to ensure that the height of the simulation cell is greater than twice the cut-off radius of the interatomic potential \cite{Smith89}. To do this 
we ensure that we have the reduced basis for the lattice which maximizes the minimum distance between faces. To remain with a compatible lattice up to a large Hencky strain we found that a combination of analytic factorisation and a numerical lattice reduction were required. 

If we have a basis matrix for the cell $\mathbf{G}$ then another equivalent basis matrix is given by $\mathbf{G}'=\mathbf{N}\mathbf{G}$
where $\mathbf{N}$ is a unimodular integer matrix. In our case the initial basis matrix $\mathbf{B}$ is a Vandermonde matrix which obeys the relation $\mathbf{B}\mathbf{D}=\mathbf{C}\mathbf{B}$, where $\mathbf{C}$ is the companion matrix
\begin{equation}
\mathbf{C} = \left( \begin{array}{ccc} 
      0 & 1 & 0 \\
      0 & 0 & 1  \\
      1 & -5 & 6
   \end{array}\right)
\end{equation} 
and $\mathbf{D}=\textrm{diag}(\phi_1,\phi_2,\phi_3)$ \cite{Brand64}. The deformation tensor for uniaxial extension is, $\boldsymbol{\Lambda}(t) = \textrm{diag}(e^{\dot \epsilon t},e^{-\dot \epsilon t/2},e^{-\dot \epsilon t/2})$. We can factor out $\mathbf{D}$ from $\boldsymbol{\Lambda}(t)$ to give, 
\begin{equation}
\boldsymbol{\Lambda}(t) = \mathbf{D}^{k_3(t)}\boldsymbol{\Delta}(t)
\end{equation}
where,
\begin{equation}
\label{eq:Delta}
\boldsymbol{\Delta}(t) = \textrm{diag}(e^{(k_1(t) - k_3(t)) ln \phi_1 +\delta_1(t)},e^{(k_2(t) - k_3(t)) ln \phi_2 +\delta_2(t)},e^{\delta_3(t)})
\end{equation}

and 
$k_1(t)=\lfloor\dot{\epsilon}t/ln\phi_1\rfloor,
  k_2(t)=-\lfloor\dot{\epsilon}t/(2ln\phi_2)\rfloor,
  k_3(t)=-\lfloor\dot{\epsilon}t/(2ln\phi_3)\rfloor,$ 
  and
  $\delta_1(t) = \dot{\epsilon} t - k_1(t)ln\phi_1  ,
  \delta_2(t) = -\dot{\epsilon} t/2 - k_2(t)ln\phi_2,
  \delta_3(t)= -\dot{\epsilon} t/2 - k_3(t)ln\phi_3$.
   Note that our choice of $\phi_1, \phi_2$ and $\phi_3$ mean that for positive $t$, $k_1(t), k_2(t)$ and $k_{3}(t)$ are positive integers. One can now write,
\begin{eqnarray}
\mathbf{B} \boldsymbol{\Lambda}(t) &=& \mathbf{B}\mathbf{D}^{k_3(t)}\boldsymbol{\Delta}(t) \nonumber \\
&=&\mathbf{C}^{k_3(t)} \mathbf{B} \boldsymbol{\Delta}(t).
\end{eqnarray}
The matrix $\mathbf{C}^{k_3(t)}$ is a unimodular integer matrix, so that the matrix $\mathbf{B}\boldsymbol{\Delta}(t)$ provides a more compact basis for the lattice. However, this is not necessarily the 
reduced basis for the lattice. A further calculation is required to obtain the fully reduced basis. For this a standard algorithm, the so called LLL algorithm, is used \cite{LLL}. 
This algorithm gives the reduced basis of a lattice having vectors with only integer components. To use the LLL algorithm we scale 
$\mathbf{B}\boldsymbol{\Delta}(t)$ 
by a very large factor $g$ and round each element to its nearest integer giving  
$g\mathbf{B}\boldsymbol{\Delta}(t)\approx\lfloor g\mathbf{B}\boldsymbol{\Delta}(t)\rceil \equiv \mathbf{M}$. In our implementation we have used 
$g=10^{160}$. Performing LLL reduction gives 
$\mathbf{M}'=\mathrm{LLL}(\mathbf{M})=\mathbf{QM}$ where 
$\mathbf{Q}=\mathbf{M}'\mathbf{M}^{-1}$ 
is a unimodular integer matrix. 
$\mathbf{Q}$ 
is applied to $\mathbf{B}\boldsymbol{\Delta}(t)$ to obtain the reduced basis,
\begin{equation}
\mathbf{B}'(t) = \mathbf{Q}\mathbf{B}\boldsymbol{\Delta}(t)
\end{equation}

In summary our algorithm for calculating $\mathbf{B}'(t)$ at each time step is as follows:
(1) Calculate $\boldsymbol{\Delta}(t)$ using Eq. (\ref{eq:Delta}) and the definitions for $k_i(t)$ and $\delta_i(t)$.
(2) Find $\mathbf{M}=\lfloor g \mathbf{B} \boldsymbol{\Delta}(t)\rceil$.
(3) Find the lattice reduced version of $\mathbf{M}$ using the LLL algorithm \textit{i.e.}
$\mathbf{M}'=\textrm{LLL}(\mathbf{M})=\mathbf{QM}$ and calculate $\mathbf{Q}=\mathbf{M}'\mathbf{M}^{-1}$.
(4) Lastly calculate $\mathbf{B}'(t)=\mathbf{QB}\boldsymbol{\Delta}(t)$.

At most time steps  $\mathbf{B}'(t+\delta t)$ is simply the deformation of $\mathbf{B}'(t)$.
However, there is a true switch of  the basis vectors on average after $\epsilon=0.17$.
The switching time is not exactly periodic as it is for Kraynik-Reinelt PBCs. 
We find numerically that up to a Hencky strain of $1000$ 
that the height has a minimum value $h_{min}\approx1.442$. It is important to note that to get to large 
Hencky strains we need to  use arbitrary precision arithmetic. The deformed lattice given by
 $\mathbf{B}\boldsymbol{\Delta}(t)$ can be rather stretched and so to have $\mathbf{B}'$ 
remain on the hyperbolic surface both  $\mathbf{B}$ and $\boldsymbol{\Delta}$ should be 
calculated with some care. We have performed the lattice calculation and the LLL reduction 
using \textsf{Sage} \cite{Sage}. For a specific strain rate, the basis for each time step is first written 
(with double precision) to a file which is then read by the molecular dynamics code during
a simulation. The basis from the file is scaled to give a simulation cell with the correct volume.
The arbitrary precision libraries used by \cite{Sage} and LLL libraries will be implemented 
in the future to avoid the need for the production of a file with a time series of basis vectors.
An article is in preparation which provides more details of the algorithm \cite{Hunt2014}.  

 It is possible to use the methods above for other three dimensional extensional flows, including biaxial extensional flow.  For the flow with the velocity gradient
$\nabla \mathbf{u} = \textrm{diag}(ln \phi_1,ln \phi_2,ln \phi_3)$ the 
lattice is reproduced at time $t=1$. We can also show that under UEF that the 
individual $x, y$ and $z$ components of the basis vectors are reproduced at incommensurate times, \textit{i.e.} they are an integer linear combination of their original values. 
This agrees with
the finding \cite{KR92} that there are no lattices which are reproduced after some time under UEF. 

\section{Simulations and results} 
To test the algorithm we have performed NEMD simulations of UEF using the SLLOD equations of motion \cite{EM90,Dai06}
\begin{eqnarray}
  \label{eq:sllod.shear}
 \dot \mathbf{r}_i & = &\frac{\mathbf{p}_i}{m_i}+ \mathbf{r}_i \cdot \nabla \mathbf{u},\\
  \dot\mathbf{p}_i & = &\mathbf{F}_i - \mathbf{p}_i  \cdot \nabla \mathbf{u} - \alpha \mathbf{p}_i .
\end{eqnarray}
Where $\alpha$ is the Nos\'{e}-Hoover thermostat satisfying 
\begin{equation}
\dot\alpha=\frac{1}{Q}\left[  \sum_{i=1}^{N}\frac{\mathbf{p}_i^2}{m_i} - N_f k_B T \right]
\end{equation}
with  a damping factor $Q=10.0$ and $N_f$ the number of degrees of freedom in the simulation.
In addition to these equations of motion we have also had to periodically zero the centre of mass momenta in a similar way done for the planar extensional flow \cite{TD00} . 
A system of $1728$ atoms was simulated with the interatomic potential of Weeks, Chandler and Anderson  $U_{WCA}(r) = 4\epsilon\left[\left(\frac{\sigma}{r}\right)^{12}-\left(\frac{\sigma}{r}\right)^{6}\right]+\epsilon $ for $r\le 2^{1/6}\sigma$ and $0$ for $r>2^{1/6}\sigma$, we use reduced units setting $\epsilon$ and $\sigma$ to unity. The systems were simulated at the Lennard-Jones triple point $\rho=0.8442$ and $T=0.722$. 

To test the new PBC algorithm we have performed simulations using the simple deforming PBCs and the new technique.
In the case of the simple deforming PBCs the initial basis vectors of the lattice are the usual cartesian basis vectors scaled to give a simulation cell with volume $V=L^3$. The time dependent vectors are given by the rows of the matrix 
$\mathbf{B}_{old}(t)=\textrm{diag}(L e^{\dot\epsilon t},L e^{-\dot\epsilon t/2}, L e^{-\dot\epsilon t/2})$.

\begin{figure}
\subfigure{
\label{fig:start_up}
\includegraphics[width=8.5cm,height=6.0cm]{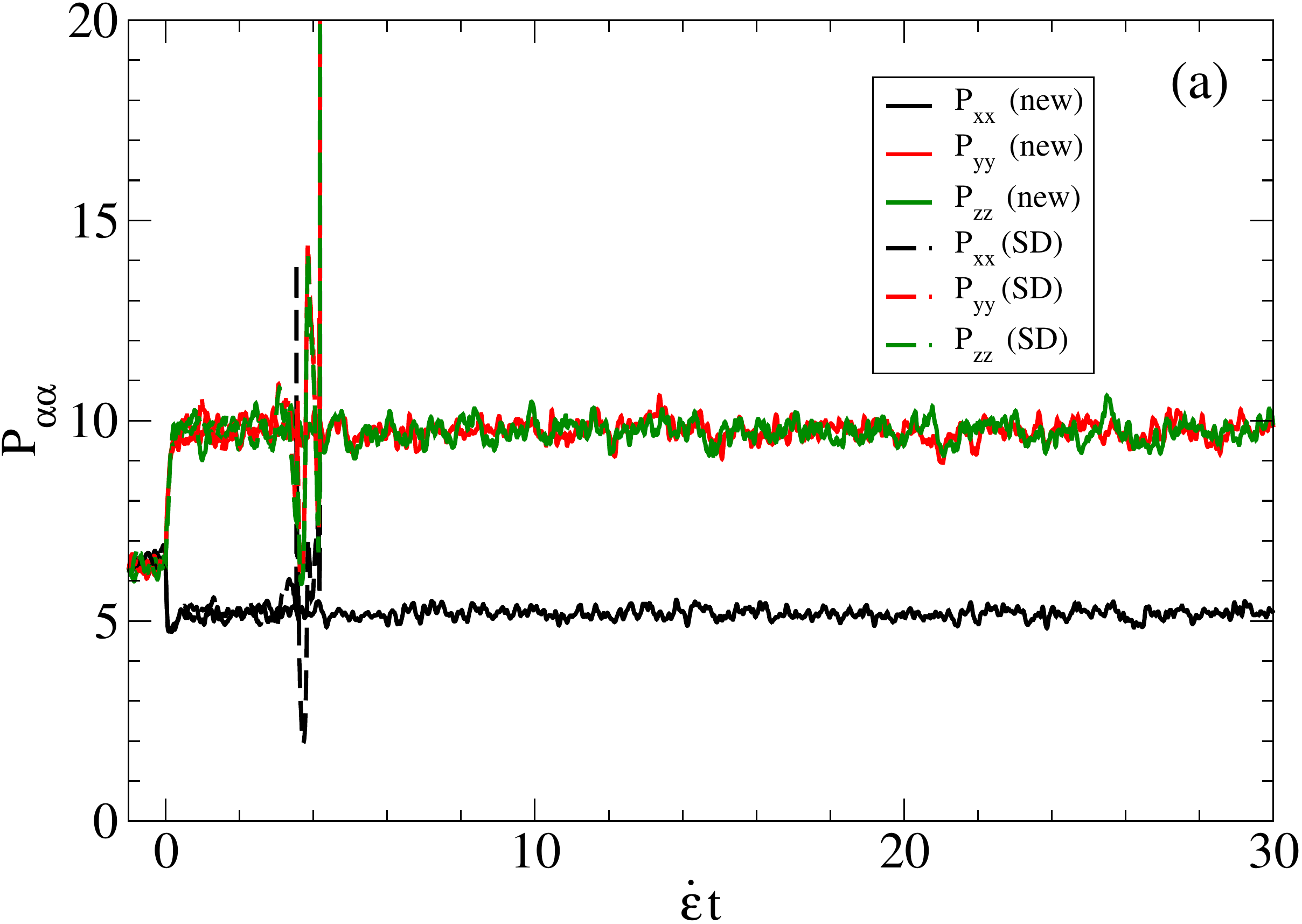} 
}
\subfigure{
\label{fig:press_ss} 
\includegraphics[width=8.5cm]{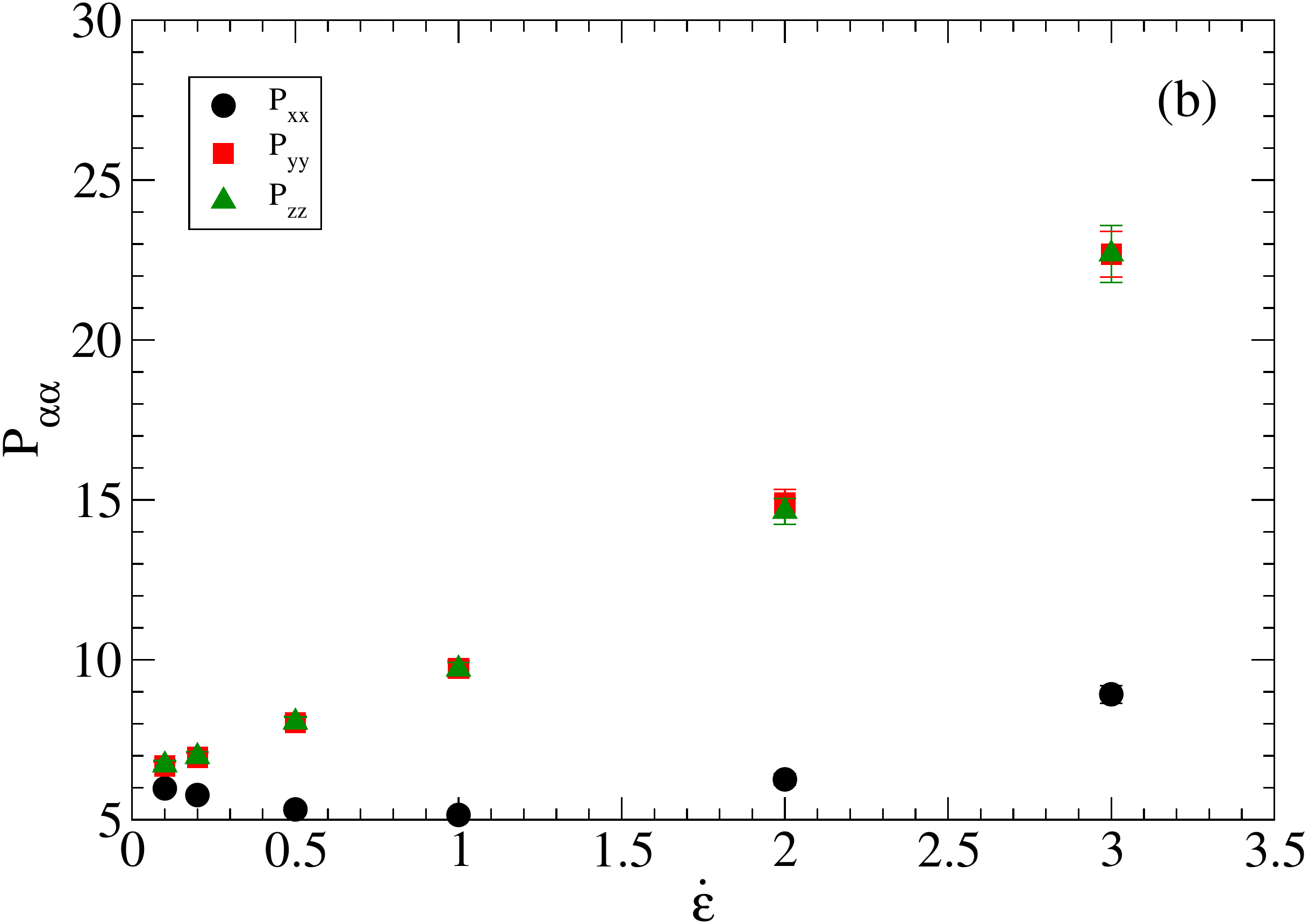} 
}
\caption{ (a) The diagonal components of the pressure tensor during start-up and steady state of uniaxial extensional flow with a Hencky strain rate $\dot\epsilon = 1.0$, using the simple deforming PBCs (dashed lines) and new PBCs (full lines). (b) The diagonal components of the pressure tensor at steady-state under a range of strain rates.}
\end{figure}

The evolution of the diagonal components of the pressure tensor \cite{EM90} of the system during start-up of UEF at a Hencky strain rate of $\dot\epsilon=1.0$ are shown in (Fig. \ref{fig:start_up}). In the period up to a Hencky strain of $3.0$ results from the old and new PBCs agree, after which the old PBCs produce results which fluctuate significantly before the simulation stops.  In contrast, using the new PBCs the simulations were kept running at a steady-state up to a Hencky strain of $30$ and could have been kept running considerably longer.

The steady state components of the pressure tensor for various values of the strain rate are given in (Fig.\ref{fig:press_ss}). These results can be compared with the results of Todd and Daivis \cite{Todd97} who calculated extensional properties through direct simulations using the simple deforming PBCs and by extrapolation of oscillatory extensional data on the same model liquid tested here.

\begin{acknowledgments}
S. Bernardi, W.K. den Otter, W.J. Briels, F. Frascoli, P.J. Daivis, A.M. Kraynik and B.D. Todd are thanked for helpful discussions. B.D. Todd, W.J. Briels and in particular S. Bernardi are thanked for comments on a draft manuscript.  This work is part of the Industrial Partnership Programme (IPP) `Bio(-related) Materials' of the \textit{`Stichting voor Fundamenteel Onderzoek der Materie FOM'}, which is supported financially by the \textit{`Nederlandse Organisatie voor Wetenschappelijk Onderzoek (NWO)'}. This IPP is co-financed by the Top Institute Food and Nutrition and the Dutch Polymer Institute.
\end{acknowledgments}

\end{document}